\documentclass[prd,preprint,superscriptaddress,showpacs,byrevtex]{revtex4}
\usepackage{epsfig}

\newcommand{\be}{\begin{equation}}
\newcommand{\ee}{\end{equation}}
\newcommand{\bea}{\begin{eqnarray}}
\newcommand{\eea}{\end{eqnarray}}

\begin{document}

\begin{flushright}
YITP-SB-07-28
\end{flushright}

\title{ Schwinger Mechanism for Gluon Pair Production in the Presence of
Arbitrary Time Dependent Chromo-Electric Field }

\author{Gouranga C. Nayak} \email{nayak@max2.physics.sunysb.edu}

\affiliation{C. N. Yang Institute for Theoretical Physics, Stony Brook
University, SUNY, Stony Brook, NY 11794-3840, USA }

\date{\today}

\begin{abstract}

We study Schwinger mechanism for gluon pair production in the presence
of arbitrary time-dependent chromo-electric background field $E^a(t)$
with arbitrary color index $a$=1,2,...8 in SU(3) by directly evaluating
the path integral. We obtain an exact expression for the probability of
non-perturbative gluon pair production per unit time per unit volume and
per unit transverse momentum $\frac{dW}{d^4x d^2p_T}$ from arbitrary
$E^a(t)$. We show that the tadpole (or single gluon) effective action
does not contribute to the non-perturbative gluon pair production rate
$\frac{dW}{d^4x d^2p_T}$.
We find that the exact result for non-perturbative gluon pair production
is independent of all the time derivatives $\frac{d^nE^a(t)}{dt^n}$
where $n=1,2,....\infty$ and has the same functional dependence on two
casimir invariants $[E^a(t)E^a(t)]$ and $[d_{abc}E^a(t)E^b(t)E^c(t)]^2$
as the constant chromo-electric field $E^a$ result with the replacement:
$E^a \rightarrow E^a(t)$. This result may be relevant to study the
production of a non-perturbative quark-gluon plasma at RHIC and LHC.

\end{abstract}
\maketitle

\newpage

\section {Introduction}

An exact non-perturbative result for electron-positron
pair production from a constant electric field was obtained
by Schwinger in 1951 by using proper time method \cite{schw}. In QCD
this result depends on two independent casimir/gauge invariants $C_1=[E^aE^a]$
and $C_2=[d_{abc}E^aE^bE^c]^2$ with color indices $a,b,c$=1,2,...8 in SU(3)
\cite{gn}. Recently we have extended this calculation to arbitrary
time dependent electric field $E(t)$ in QED \cite{nayak1}
and to arbitrary time dependent
chromo-electric field $E^a(t)$ in QCD for quark-antiquark case
\cite{nayak2}. This result relies crucially on the validity of the shift
conjecture, which has not yet been established. In this paper we will extend this
calculation to study non-perturbative
gluon pair production from arbitrary $E^a(t)$. Unlike constant field $E^a$
case \cite{gn}, we encounter a non-vanishing single gluon (or tadpole) term
in the presence of time dependent chromo-electric
field $E^a(t)$. However, we show that the non-perturbative
tadpole effective action $\frac{dS_{\rm tad}}{d^4xd^2p_T}$
contains $\delta^{(2)}(\vec{p}_T)$ distribution and hence, for
any non-vanishing transverse momentum, it does not contribute to
non-perturbative
gluon pair production rate $\frac{dW}{d^4x d^2p_T}$ via Schwinger mechanism.
This result may be relevant to study the production of a non-perturbative
quark-gluon plasma at RHIC and LHC \cite{gn,qgp1,qgp2,qgp3}.

We obtain the following exact non-perturbative
result for the probability of gluon (pair)
production per unit time, per unit volume and per unit transverse momentum
from an arbitrary time dependent chromo-electric field $E^a(t)$ with
arbitrary color index $a$=1,2,...8 in SU(3):
\bea
\frac{dW_{g (\bar g)}}{dt d^3x d^2p_T}~
=~\frac{1}{4 \pi^3} ~~ \sum_{j=1}^3 ~
~|g\Lambda_j(t)|~{\rm ln}[1~+~e^{-\frac{ \pi p_T^2}{|g\Lambda_j(t)|}}].
\label{1}
\eea
In the above equation
\bea
\Lambda_1^2(t)=\frac{C_1(t)}{2}[1-{\rm cos}\theta(t)]; ~~~
\Lambda_{2,3}^2(t)=\frac{C_1(t)}{2}[1+{\rm cos}(\frac{\pi}{3} \pm \theta(t))];~~~cos^3\theta(t)=-1+6 C_2(t)/C_1^3(t) \nonumber \\
\label{lm}
\eea
where
\bea
~~~~~~~~~~~~C_1(t)=[E^a(t)E^a(t)];~~~~~~~~~~~~~~{\rm and}~~~~~~~~~~~ C_2(t)=[d_{abc}E^a(t)E^b(t)E^c(t)]^2
\label{cas}
\eea
are two independent time-dependent casimir/gauge invariants in SU(3).

This result has the remarkable feature that it is independent of all the time
derivatives $\frac{d^nE^a(t)}{dt^n}$ and
has the same functional form as the constant chromo-electric field $E^a$
result \cite{nayak1} with: $E^a \rightarrow E^a(t)$.

We will present a derivation of eq. (\ref{1}) in this paper.

\section { Schwinger Mechanism in QCD in the presence of Arbitrary
Time-Dependent Chromo-Electric Field }

In the background field method of QCD \cite{thooft,abbott} the
gauge field is the sum of classical chromo-field $A_\mu^a$ and the quantum
gluon field $Q_\mu^a$. The non-abelian field tensor becomes
\bea
F_{\mu \nu}^a[A+Q]= \partial_\mu (A_\nu^a+Q_\nu^a) -\partial_\nu (A_\mu^a+Q_\mu^a) +gf^{abc} (A_\mu^b+Q_\mu^b) (A_\nu^c+Q_\nu^c).
\eea
The gauge field Lagrangian density is
\bea
{\cal L}_{gl}~=~-\frac{1}{4} F_{\mu \nu}^a[A+Q] F^{\mu \nu a}[A+Q]
~-\frac{1}{2\alpha} [D_\mu[A]Q^{\mu a}]^2
\label{total1}
\eea
where the second term in the right hand side is the gauge fixing term
\cite{thooft,abbott}. The covariant derivative is given by
\bea
D_\mu^{ab}[A]~=~\delta^{ab} \partial_\mu~+~gf^{abc}A_\mu^c.
\eea
Keeping terms up to quadratic in $Q$ field (for gluon pair production) and using
Feynman-t' Hooft gauge ($\alpha$=1) we find from eq. (\ref{total1})
\bea
&&~\int d^4x {\cal L}= \frac{1}{2} \int d^4x~[-(D_\mu[A]Q_\nu^a)F^{\mu \nu a}[A] +
Q^{\mu a} M^{ab}_{\mu \nu}[A] Q^{\nu b} ] \nonumber \\
&& = \frac{1}{2} \int d^4x~[(D_\mu[A]F^{\mu \nu a}[A]) Q_\nu^a +
Q^{\mu a} M^{ab}_{\mu \nu}[A] Q^{\nu b} ]
\label{full}
\eea
where
\bea
M^{ab}_{\mu \nu}[A]~=~
g_{\mu \nu} [D^2(A)]^{ab}~-~2gf^{abc}F_{\mu \nu}^c[A]
\label{mab}
\eea
with $g_{\mu \nu}=(1,-1,-1,-1)$.

The vacuum-to-vacuum transition amplitude for gluon (we will consider the
ghost later in the derivation) in the presence of classical chromo-field
$A_\mu^a$ is given by
\bea
<0|0>^A=\frac{Z[A]}{Z[0]}=
\frac{\int [dQ]e^{i\int d^4x [Q^{\mu a}M_{\mu \nu}^{ab}[A]Q^{\nu b}+(D_\mu [A] F^{\mu \nu a}[A])Q_\nu^a]
}}{\int [dQ] e^{i\int d^4xQ^{\mu a}M_{\mu \nu}^{ab}[0]Q^{\nu b}}}.
\label{vact}
\eea

We choose the arbitrary time-dependent chromo-electric field
$E^a(t)$ to be along the $z-$axis
(the beam direction) and work in the choice $A_3^a=0$
\cite{nayak2} so that
\bea
A_\mu^a(x) = -\delta_{\mu 0} E^a(t) z.
\label{gauge}
\eea
The color indices $a$=1,2,...8 are arbitrary. From eq. (\ref{gauge}) we find
\bea
D_\mu [A] F^{\mu \nu a}[A] =\frac{dE^a(t)}{dt} ~\delta^\nu_3.
\label{sing}
\eea
Hence the single gluon term
$(D_\mu [A] F^{\mu \nu a}[A])Q_\nu^a$ in eq. (\ref{full}) was absent
for constant chromo-electric field $E^a$ case in \cite{nayak1}.
However, in the presence of time-dependent chromo-electric field $E^a(t)$
this single gluon term in eq. (\ref{full}) is not zero.
Hence the path integration in eq. (\ref{vact}) becomes more complicated
due to the presence of this single gluon term.

To evaluate the path integration in eq. (\ref{vact}) we proceed
as follows. We write
\bea
M_{\mu \nu}^{ab}(x,x')=\delta^{(4)}(x-x') M_{\mu \nu}^{ab}(x')
\label{m1}
\eea
where we denote
$M_{\mu \nu}^{ab}(x) =M_{\mu \nu}^{ab}[A](x)$ which is given by eq. (\ref{mab}).
The Green's function $G_{\mu \nu}^{ab}(x,x') (=[{M^{-1}}]_{\mu \nu}^{ab}(x,x'))$ is
given by
\bea
\int d^4x'' M^{\mu \lambda, ac}(x,x'') G_{\lambda \nu}^{cb}(x'',x') = \delta^{ab} g^\mu_\nu \delta^{(4)}(x-x').
\label{green}
\eea

We change the variable
\bea
{Q}_\mu^a(x)={Q'}_\mu^a(x)-\frac{1}{2}~\int d^4x' G_{\mu \nu}^{ab}(x,x')D_\lambda (x')F^{\lambda \nu b}(x')
\label{qp}
\eea
where we denote $D_\mu^{ab}(x)=D_\mu^{ab}[A](x)$ and
$F_{\mu \nu}^a(x)= F_{\mu \nu}^a[A](x)$.
Under this change of variable $[dQ]=[dQ']$.
Using eqs. (\ref{qp}), (\ref{green}) and (\ref{m1}) we find
\bea
&& [(D_\mu[A]F^{\mu \nu a}[A]) Q_\nu^a +
Q^{\mu a} M^{ab}_{\mu \nu}[A] Q^{\nu b} ]  \nonumber \\
&& = -\frac{1}{2}~\int d^4x' D_\mu(x) F^{\mu \lambda a}(x)G_{\mu \nu}^{ab}(x,x')
D_\sigma (x')F^{\sigma \nu b}(x') +{Q'}^{\mu a}M_{\mu \nu}^{ab}[A]{Q'}^{\nu b}
\label{full1}
\eea
Hence we find from eq. (\ref{vact})
\bea
&& <0|0>^A=\frac{Z[A]}{Z[0]}=
\exp[-\frac{i}{2}\int d^4x \int d^4x' D_\mu(x) F^{\mu \lambda a}(x)G_{\mu \nu}^{ab}(x,x')
D_\sigma (x')F^{\sigma \nu b}(x')] \nonumber \\
&&~\times ~
\frac{\int [dQ']e^{i\int d^4x {Q'}^{\mu a}M_{\mu \nu}^{ab}[A]{Q'}^{\nu b}}}{\int [dQ] e^{i\int d^4xQ^{\mu a}M_{\mu \nu}^{ab}[0]Q^{\nu b}}}
 =e^{-i S_{\rm tad}} \times \frac{{\rm Det^{-1/2}}M_{\mu \nu}^{ab}[A]}{{\rm Det^{-1/2}}M_{\mu \nu}^{ab}[0]}=
e^{-i S_{\rm tad}} \times e^{iS^{(1)}}
\label{vact1}
\eea
where
\bea
S_{\rm tad} = \frac{1}{2}~\int d^4x \int d^4x' D_\mu(x) F^{\mu \lambda a}(x) G_{\lambda}^{ \nu ab}(x,x')
D^\sigma (x') F_{\sigma \nu b}(x')
\label{tadd}
\eea
is the tadpole (or single gluon) effective action and
\bea
S^{(1)}=-i {\rm ln}[ \frac{{\rm Det^{-1/2}}M_{\mu \nu}^{ab}[A]}{{\rm Det^{-1/2}}M_{\mu \nu}^{ab}[0]}]
\label{og}
\eea
is the one loop (or gluon pair) effective action.

\subsection { Tadpole (or Single Gluon) Effective Action in Arbitrary $E^a(t)$ }

For an operator $M$ we will use Schwinger's notation \cite{schw} for the
Green's function
\bea
G(x,x') = <x|\frac{1}{M}|x'>=<x|\int_0^\infty ds~ e^{-sM }  |x'>.
\eea
The Green's function $G_{\mu \nu}^{ab}(x,x')$
for the operator $M_\mu^{ \nu ab}[A]$ becomes
\bea
G_\mu^{ \nu ab}(x,x') = <x| \int_0^\infty ds~ e^{-sM_\mu^{ \nu ab}[A]}|x'>.
\label{gr1}
\eea
Using eq. (\ref{gauge}) in (\ref{mab}) we find
\bea
M_{\mu}^{ \nu ab}[A]=
M_{\mu \lambda}^{ ab}[A] g^{\lambda \nu}=
-\delta_\mu^\nu[(\delta^{ab}{\hat p}_0-igf^{abc}E^c(t) z)^2 -\delta^{ab}{\hat p}_z^2 -\delta^{ab}{\hat p_T}^2]-~2gf^{abc}E^c(t) {\hat F}_\mu^\nu
\label{mabn}
\eea
where
\bea
{\hat F}_\mu^\nu~=
\left [
\begin{array}{cccc}
0 & 0 & 0 & 1 \\
0 & 0 & 0 & 0 \\
0 & 0 & 0 & 0 \\
1 & 0 & 0 & 0
\end{array}
\right ].
\label{mat}
\eea
By using eq. (\ref{mabn}) in (\ref{gr1}) we find
\bea
G_\mu^{\nu ab}(x,x') = <x| \int_0^\infty ds~ e^{-s(
\delta_\mu^\nu[-(\delta^{ab}{\hat p}_0-igf^{abc}E^c(t) z)^2 +\delta^{ab}{\hat p}_z^2 +\delta^{ab}{\hat p_T}^2]-~2gf^{abc}E^c(t) {\hat F}_\mu^\nu
}|x'>.
\label{gr2}
\eea
We write eq. (\ref{gr2}) in the Lorentz and color matrix notation as follows
\bea
G_\mu^{\nu, ab}(x,x')
= \int_0^\infty ds[<x| \int_0^\infty ds ~e^{-s( -({\hat p}_0 -
g\Lambda(t) z)^2 +\hat{p}_z^2 +\hat{p}_T^2+~2ig \Lambda(t){\hat F} )}  |x'>]_\mu^{\nu ab}
\label{gr}
\eea
where
\bea
\Lambda^{ab}(t)=if^{abc}E^c(t).
\label{lamb}
\eea
Using eqs. (\ref{gr}) and (\ref{sing})
in (\ref{tadd}) we find the tadpole effective action
\bea
&& S_{\rm tad}
=\frac{1}{2}~ \int d^4x d^4x' \int_0^\infty ds ~\frac{dE^a(t)}{dt}
[<x|  e^{-s(- ({\hat p}_0 -i
g\Lambda(t) z)^2 +\hat{p}_z^2 +\hat{p}_T^2+~2ig \Lambda(t){\hat F}  )}  |x'>]
_3^{3 ab}
\frac{dE^b(t')}{dt'} \nonumber \\
&& = \frac{1}{2}~\int d^2x_T ~d^2x'_T~  dt~ dz~ dt'~ dz'
\int_0^\infty ds \nonumber \\
&& \frac{dE^a(t)}{dt} [<t|<z|<x_T|  e^{-s(- ({\hat p}_0 -i
g\Lambda(t) z)^2 +\hat{p}_z^2 +\hat{p}_T^2+~2ig \Lambda(t){\hat F}  )}
|x'_T>|z'>|t'>] _3^{3 ab}
\frac{dE^b(t')}{dt'}.
\label{tad2}
\eea
Inserting complete set of $|p_T>$ states (by using $\int d^2p_T |p_T><p_T|=1$)
we find
\bea
&& S_{\rm tad} = \frac{1}{2}~\int d^2x_T ~d^2x'_T~  dt~ dz~ dt'~ dz' ~ d^2p_T~\int_0^\infty ds~ \frac{dE^a(t)}{dt}
[<x_T|p_T>  \nonumber \\
&& <t|<z|~ e^{-s(- ({\hat p}_0 -i
g\Lambda(t) z)^2 +\hat{p}_z^2 +p_T^2+~2ig \Lambda(t){\hat F} )} |t'>|z'> <p_T|x'_T>]_3^{3 ab}
\frac{dE^b(t')}{dt'}.
\label{tad1}
\eea
Using $<q|p>=\frac{1}{\sqrt{2\pi}} e^{iqp}$ we obtain
\bea
&& S_{\rm tad} = \frac{1}{2(2\pi)^2} \int d^2x_T ~d^2x'_T ~ dt ~dz ~dt' ~dz'~  d^2p_T~ \int_0^\infty ds~\frac{dE^a(t)}{dt}
[e^{ix_T \cdot p_T}  \nonumber \\
&&~ <t|<z|e^{-s(- ({\hat p}_0 -ig\Lambda(t) z)^2 +\hat{p}_z^2 +p_T^2+~2ig \Lambda(t){\hat F})}
|z'>|t'> e^{-ix'_T \cdot p_T}]_3^{3 ab} \frac{dE^b(t')}{dt'}.
\label{tad3}
\eea
Integrating over $x'_T$ (by using $\int d^2x'_T e^{-ix'_T \cdot p_T} = (2 \pi)^2 ~\delta^{(2)}(\vec{p}_T))$
we find
\bea
&& \frac{dS_{\rm tad}}{d^4xd^2p_T} =\frac{1}{2}~\delta^{(2)}(\vec{p}_T)~
e^{ix_T \cdot p_T} ~
\int dt'~\int dz'  ~\int_0^\infty ds~\nonumber \\
&& \frac{dE^a(t)}{dt} [<t|<z|e^{-s(- ({\hat p}_0 -
ig\Lambda(t) z)^2 +\hat{p}_z^2 +p_T^2 +~2ig \Lambda(t){\hat F} )} |z'>|t'>]_3^{3 ab} \frac{dE^b(t')}{dt'}.
\nonumber \\
\label{tad4}
\eea
Since the above equation contains Dirac-delta function
$\delta^{(2)}(\vec{p}_T)$ we find (for any non-vanishing $p_T$)
\bea
\frac{dS_{\rm tad}}{d^4xd^2p_T}=0.
\label{pt}
\eea
Hence the tadpole (or single gluon) effective action
does not contribute to the exact result ($\frac{dW}{d^4xd^2p_T}$) for the probability
of gluon (pair) production per unit time per
unit volume per unit transverse momentum
from an arbitrary $E^a(t)$ via Schwinger mechanism.

Now we evaluate the one-loop effective action for gluon and ghost in the
presence of arbitrary $E^a(t)$ in the following.

\subsection { One Loop (or Gluon Pair) Effective Action in Arbitrary $E^a(t)$ }

The one loop (or gluon pair) effective action eq. (\ref{og}) can be written as
\bea
&& S^{(1)}=-i {\rm ln}[ \frac{{\rm Det^{-1/2}}M_{\mu \nu}^{ab}[A]}{{\rm Det^{-1/2}}M_{\mu \nu}^{ab}[0]}]=\frac{i}{2}{\rm Tr}[ {\rm ln} M_\mu^{ \nu ab}[A] -{\rm ln} M_\mu^{ \nu ab}[0]] \nonumber \\
&&
=\frac{i}{2} {\rm Tr} \int_0^\infty \frac{ds}{s} [ e^{is(M_\mu^{ \nu ab}[A] +i\epsilon)}  -e^{is( M_\mu^{ \nu ab}[0] +i\epsilon)}].~~~
\label{vc3}
\eea
The trace ${\rm Tr}$ is given by
\bea
{\rm Tr} {\cal O} = {\rm tr}_{\rm Lorentz } {\rm tr}_{\rm color}\int d^4x <x|{\cal O} |x>.
\eea
Using eq. (\ref{mabn}) in (\ref{vc3}) and writing in matrix notations we find
\bea
S^{(1)}=\frac{i}{2} {\rm tr}_{\rm Lorentz } {\rm tr}_{\rm color} \int d^4x
 \int_0^\infty \frac{ds}{s}
[ <x|e^{-is(({\hat p}_0-g\Lambda(t) z)^2 -{\hat p}_z^2 -{\hat p_T}^2-~2ig\Lambda(t) {\hat F}
 -i\epsilon)}  -e^{-is({\hat p}^2 -i\epsilon)}|x>]_{\mu}^{\nu, ab} \nonumber \\
\label{vac4}
\eea
where the Lorentz matrix ${\hat F}_\mu^\nu$ and the color matrix
$\Lambda^{ab}$ matrices are given by eqs. (\ref{mat}) and (\ref{lamb})
respectively.

Using the eigen values
\bea
{\hat F}_{\rm eigenvalues} =(\lambda_1, \lambda_2, \lambda_3, \lambda_4) = (1,0,-1,0).
\label{eigend}
\eea
we perform the Lorentz trace and find
\bea
&& S^{(1)} =\frac{i}{2} \sum_{l=1}^4 {\rm tr}_{\rm color}~[\int_0^\infty \frac{ds}{s}
\int dt <t| \int dx <x| \int dy <y| \int dz <z| \nonumber \\
&&
e^{-is(({\hat p}_0-g\Lambda(t) z)^2 -{\hat p}_z^2 -{\hat p_T}^2-~2i\lambda_l g\Lambda(t)
 -i\epsilon)}  -e^{-is({\hat p}^2 -i\epsilon)}
 |z> |y> |x> |t>]^{ab}.
\eea
Inserting complete set of $|p_T>$ states (using $\int d^2p_T~ |p_T><p_T|=1$)
we find from the above equation
\bea
&& S^{(1)}=\frac{i}{2(2\pi)^2}  \sum_{l=1}^4
{\rm tr_{color}}[
\int_0^\infty\frac{ds}{s} \int d^2x_T\int d^2p_T
e^{is(p_T^2+i\epsilon)} \nonumber \\
&&~[ \int_{-\infty}^{+\infty} dt <t| \int_{-\infty}^{+\infty} dz <z|
 e^{-is[(i \frac{d}{dt} -g\Lambda(t) z)^2-\hat{p}_z^2 -2ig\lambda_l \Lambda(t)]}
|z>|t> - \int dt  \int dz  \frac{1}{4\pi s}]
]^{ab}
\nonumber \\
\label{2ja}
\eea
where we have used the normalization $<q|p>=\frac{1}{\sqrt{2\pi}} e^{iqp}$.
At this stage we use the shift theorem \cite{nayak4} and find
\bea
&& S^{(1)}=\frac{i}{2(2\pi)^2}  \sum_{l=1}^4
{\rm tr_{color}}[
\int_0^\infty\frac{ds}{s} \int d^2x_T\int d^2p_T
e^{is(p_T^2+i\epsilon)} [ \int_{-\infty}^{+\infty} dt <t| \int_{-\infty}^{+\infty} dz \nonumber \\
&& <z+\frac{i}{g\Lambda(t)}\frac{d}{dt}|
 e^{-is[g^2\Lambda^2(t) z^2-\hat{p}_z^2 -2ig\lambda_l \Lambda(t)]} |z+\frac{i}{g\Lambda(t)}\frac{d}{dt}>|t> - \int dt  \int dz  \frac{1}{4\pi s}]
]^{ab}
\nonumber \\
\label{12a}
\eea
where the $z$ integration must be performed from $-\infty$ to $+\infty$
for the shift theorem to be applicable.

Note that a state vector $|z+\frac{i}{a(t)}\frac{d}{dt}>$ which contains
derivative operator is not familiar in physics.
However, the state vector $|z+\frac{i}{a(t)}\frac{d}{dt}>$
contains the derivative $\frac{d}{dt}$ not $\frac{d}{dz}$. Hence the
state vector is defined in the
$z$-space with $\frac{d}{dt}$ acting as a c-number shift
in $z$. To see how one operates with such state vector we find
\bea
<z+\frac{i}{a(t)}\frac{d}{dt}| p_z> f(t) =
\frac{1}{\sqrt{2\pi}} e^{i(z+\frac{i}{a(t)}\frac{d}{dt}) p_z} f(t) =
\frac{1}{\sqrt{2\pi}} e^{izp_z}e^{-\frac{p_z}{a(t)}\frac{d}{dt}} f(t).
\eea

Inserting complete sets of $|p_z>$ states (using $\int dp_z ~|p_z><p_z|=1$)
in eq. (\ref{12a}) we find
\bea
&& S^{(1)}
=\frac{i}{2(2\pi)^2} \sum_{l=1}^4
\int_0^\infty \frac{ds}{s} \int d^2x_T\int d^2p_T
e^{is(p_T^2+i\epsilon)} [ F_l(s) - \int dt \int dz~ \frac{8}{4 \pi s}]
\label{12}
\eea
where
\bea
&& F_l(s)=\frac{1}{(2\pi)}
{\rm tr_{color}}~[
\int_{-\infty}^{+\infty} dt <t|\int dp_z\int dp'_z
\int_{-\infty}^{+\infty} dz~ e^{izp_z}e^{-\frac{1}{g\Lambda(t)}\frac{d}{dt}p_z} \nonumber \\
&& <p_z| e^{is[-g^2\Lambda^2(t) z^2+\hat{p}_z^2+2i\lambda_l g\Lambda(t)]}|p'_z>
e^{\frac{1}{g\Lambda(t)}\frac{d}{dt} p'_z} e^{-izp'_z}|t> ]^{ab}.
\label{gn}
\eea
It can be seen that the exponential $e^{-\frac{1}{g\Lambda(t)}\frac{d}{dt}p_z}$
contains the derivative $\frac{d}{dt}$ which operates on
$<p_z|e^{is[-g^2\Lambda^2(t) z^2+\hat{p}_z^2-i2\lambda_l g\Lambda(t)]}|p'_z>$
hence we can not move $e^{-\frac{1}{g\Lambda(t)}\frac{d}{dt}p_z}$ to right.
We insert more complete sets of states to find
\bea
&& F_l(s) =\frac{1}{(2\pi)}
{\rm tr_{color}}~[
\int_{-\infty}^{+\infty} dt \int dt'  \int dt'' \int dz'  \int dz''
\int_{-\infty}^{+\infty} dz \int dp_0 \int dp'_0 \int dp''_0 \int dp'''_0 \nonumber \\
&& \int dp_z\int dp'_z
<t|p_0> e^{ip_z}<p_0e^{-\frac{1}{g\Lambda(t)}\frac{d}{dt}p_z}|p'_0><p'_0|t'>
<t'|<p_z|z'><z'| \nonumber \\
&&e^{is[-g^2\Lambda^2(t) z^2+\hat{p}_z^2+2i\lambda_l g\Lambda(t)]}|z''><z''|p'_z>
|t''><t'' |p''_0><p''_0|e^{\frac{1}{g\Lambda(t)}\frac{d}{dt} p'_z} |p'''_0> e^{-izp'_z}<p'''_0|t> ]^{ab} \nonumber \\
&& =\frac{1}{(2\pi)^4} {\rm tr_{color}}~[
\int_{-\infty}^{+\infty} dt \int dt'  \int dz'  \int dz''
\int_{-\infty}^{+\infty} dz \int dp_0 \int dp'_0 \int dp''_0 \int dp'''_0 \nonumber \\
&& \int dp_z\int dp'_z
e^{itp_0} e^{ip_z}<p_0e^{-\frac{1}{g\Lambda(t)}\frac{d}{dt}p_z}|p'_0>
e^{-it'p'_0} e^{iz'p_z} \nonumber \\
&& <z'|e^{is[-g^2\Lambda^2(t') z^2+\hat{p}_z^2+2i\lambda_l g\Lambda(t')]}|z''>
e^{iz''p'_z}
e^{it'p''_0}<p''_0|e^{\frac{1}{g\Lambda(t)}\frac{d}{dt} p'_z} |p'''_0> e^{-izp'_z}e^{-itp'''_0} ]^{ab}.
\label{gn151}
\eea
It can be seen that all the expressions in the above equation are independent
of $t$ except $e^{it(p_0-p'''_0)}$. This can be seen as follows
\bea
&& <p_0|f(t)\frac{d}{dt}|p'_0>= \int dt' \int dt'' \int dp''''_0 <p_0|t'> <t'|f(t)|t''><t''|p''''_0><p''''_0|\frac{d}{dt}|p'_0> \nonumber \\
&&= \int dt' \int dt'' \int dp''''_0 e^{-it' p_0}~ \delta(t'-t'') f(t'')
e^{it'' p''''_0}~
ip'_0~ \delta(p''''_0-p'_0)=ip'_0 \int dt'  ~e^{-it'(p_0-p'_0)}f(t') \nonumber \\
\label{indt}
\eea
which is independent of $t$ and $\frac{d}{dt}$.
Hence by using the cyclic property of trace we can take the matrix
$[<p''_0|e^{\frac{1}{g\Lambda(t)}\frac{d}{dt} p_z} |p'''_0>]^{ab}$ to the left.
The $t$ integration is now easy
($\int_{-\infty}^{+\infty} dt e^{it(p_0-p'''_0)}=2\pi \delta(p_0-p'''_0)$)
which gives
\bea
&& F_l(s) =\frac{1}{(2\pi)^3}
{\rm tr_{color}}~[ \int dt'  \int dz'  \int dz''
\int_{-\infty}^{+\infty} dz \int dp_0 \int dp'_0 \int dp''_0  \int dp_z\int dp'_z
e^{itp_0} e^{izp_z} \nonumber \\
&& <p''_0|e^{\frac{1}{g\Lambda(t)}\frac{d}{dt} p'_z} |p_0><p_0e^{-\frac{1}{g\Lambda(t)}\frac{d}{dt}p_z}|p'_0>e^{-iz' p_z} e^{-it'p'_0}
<z'| e^{is[-g^2\Lambda^2(t') z^2+\hat{p}_z^2+2i\lambda_l g\Lambda(t')]}|z''> \nonumber \\
&&e^{it' p''_0} e^{iz''p'_z} e^{-izp'_z}
]^{ab}.
\label{gn51}
\eea
As advertised earlier we must integrate over $z$ from $-\infty$ to $+\infty$
for the shift theorem to be applicable \cite{nayak4}. The matrix element
$<z'| e^{is[-g^2\Lambda^2(t) z^2+\hat{p}_z^2+i2\lambda_l g\Lambda(t)]}|z''>$
is independent of $z$ variable (it depends on $z'$ and $z''$ variables).
Hence we can perform the $z$ integration easily by using
$\int_{-\infty}^{+\infty} dz e^{iz(p_z-p'_z)} = 2\pi \delta(p_z-p'_z)$ to find
\bea
&& F_l(s) =\frac{1}{(2\pi)^2}
{\rm tr_{color}}~[ \int dt'  \int dz'  \int dz''  \int dp_0 \int dp'_0 \int dp''_0 \int dp_z
e^{itp_0}  \nonumber \\
&& <p''_0|e^{\frac{1}{g\Lambda(t)}\frac{d}{dt} p_z} |p_0><p_0e^{-\frac{1}{g\Lambda(t)}\frac{d}{dt}p_z}|p'_0>e^{-iz' p_z}
e^{-ip'_0t'} <z'| e^{is[-g^2\Lambda^2(t') z^2+\hat{p}_z^2+2i\lambda_l g\Lambda(t')]}|z''> \nonumber \\
&& e^{p''_0t'}e^{iz''p_z}
]^{ab}.
\label{gn51b}
\eea
Using the completeness relation $\int dp_0 |p_0><p_0|=1$ we obtain
\bea
&& F_l(s) =\frac{1}{(2\pi)^2}
{\rm tr_{color}}~[ \int dt'  \int dz'  \int dz''  \int dp'_0
\int dp_z  \nonumber \\
&& e^{-iz' p_z} <z'| e^{is[-g^2\Lambda^2(t') z^2+\hat{p}_z^2+2i\lambda_l g\Lambda(t')]}|z''>  e^{iz''p_z}
]^{ab}.
\label{gnn51f}
\eea
Since $<z'| e^{is[-g^2\Lambda^2(t') z^2+\hat{p}_z^2+2i\lambda_l g\Lambda(t')]}|z''>$ is independent of $z$
variable (it depends on $z'$ and $z''$ variables)
we can integrate over $p_z$. We find (by using
$\int dp_z e^{i(z''-z')p_z} =(2 \pi) \delta(z' - z'')$)
\bea
 F_l(s) =\frac{1}{(2\pi)}
{\rm tr_{color}}~[ \int dt  \int dp_0  \int dz'
<z'| e^{is[-g^2\Lambda^2(t) z^2+\hat{p}_z^2+2i\lambda_l g\Lambda(t)]}|z'>
]^{ab}.
\label{gn51f}
\eea
The Lorentz force equation in color space (in the adjoint representation of SU(3)) is
given by $\delta^{ab} dp_\mu =
gT^c_{ab}F^c_{\mu \nu}dx^\nu =i
gf^{abc}F^c_{\mu \nu}dx^\nu$.
When the chromo-electric field is along the $z$-axis (eq. (\ref{gauge})) this becomes
$\delta^{ab} dp_0 = igf^{abc}E^c(t) dz = g\Lambda^{ab}(t) dz$.
Using this in (\ref{gn51f}) and using the eigen values of the color
matrix $\Lambda^{ab}(t)$ \cite{gn} from eq. (\ref{lm}) we find
\bea
 F_l(s)=\frac{1}{(2\pi)} \sum_{j=1}^6 \int dt ~g\Lambda_j(t) \int dz \int dz'
 <z'| e^{is[-g^2\Lambda_j^2(t) z^2+\hat{p}_z^2+2i\lambda_l g\Lambda_j(t)]}|z'>.
\label{gn51i}
\eea
The above equation boils down to an usual
harmonic oscillator, $\omega^2(t) z'^2+\hat{p}_{z'}^2$, with the
constant frequency $\omega$ replaced
by time dependent frequency $\omega(t)$.
The normalized wave function is given by \cite{nayak2}
\bea
\int dz' |<z'|n_t>|^2=1.
\label{norm}
\eea
Inserting complete set of harmonic oscillator states
(by using $\sum_n |n_t><n_t|=1$) in eq. (\ref{gn51i}) we find
\bea
 && F_l(s)=\frac{1}{(2\pi)^2} \sum_n \sum_{j=1}^6 \int dt ~ g\Lambda_j(t) \int dz \int dz'
<z'|n_t> e^{-sg\Lambda_j(t)(2n+1)+2\lambda_l g\Lambda_j(t)} \nonumber \\
&& <n_t|z'>
=\frac{1}{(2\pi)} \sum_n \sum_{j=1}^6 \int dt ~ g\Lambda_j(t) \int dz \int dz'
|<z'|n_t>|^2 e^{-sg\Lambda_j(t)(2n+1)+2\lambda_l g\Lambda_j(t)} \nonumber \\
&& =\frac{1}{(2\pi)} \sum_{j=1}^6
\int dt \int dz~g \Lambda_j(t)
\frac{e^{sg2\lambda_l \Lambda_j(t)}}{2 {\rm sinh(sg\Lambda_j(t))}}
\label{gn51m}
\eea
where we have used eq. (\ref{norm}).
Using this expression of $F_l(s)$ in eq. (\ref{12}) and
summing over $l$ (by using the
eigen values of the Dirac matrix from eq. (\ref{eigend})) we find
\bea
S^{(1)}= \frac{i}{16 \pi^3} \sum_{j=1}^6
\int_0^\infty \frac{ds}{s} \int d^4x \int d^2p_T
e^{is(p_T^2+i\epsilon)} [g\Lambda_j(t)~
\frac{[1+{\rm cosh}(2sg\Lambda_j(t))]}{{\rm sinh}(sg\Lambda_j(t))}
-\frac{2}{s}].
\label{12ff}
\eea

\subsection { One Loop Effective Action for Ghost in arbitrary $E^a(t)$ }

Now we discuss the ghost contributions.
The ghost Lagrangian density due to the gauge fixing term in eq. (\ref{total1})
is given by \cite{thooft,abbott}
\bea
{\cal L}_{\rm gh}= {\chi^a}^\dagger [D_\mu[A] D^\mu[A+Q]]^{ab} \chi^b
= {\chi^a}^\dagger K^{ab}[A,Q] \chi^b
\label{lghost}
\eea
where $\chi^a$ is the ghost field.
The vacuum-to-vacuum transition amplitude for ghost (anti-ghost)
in the presence of $A_\mu^a(x)$ is given by:
\bea
<0|0>^A=\frac{Z[A]}{Z[0]}=
\frac{\int [d\chi^\dagger] [d\chi] e^{i\int d^4x \chi^{\dagger a}K^{ab}[A]\chi^{ b}}}{\int [d\chi^\dagger] [d\chi] e^{i\int d^4x \chi^{\dagger a}K^{ab}[0]\chi^{ b}}}=
\frac{{\rm Det}K^{ab}[A]}{{\rm Det}K^{ab}[0]}
=e^{iS^{(1)}_{\rm gh}}.
\label{vach1}
\eea
This gives
\bea
S^{(1)}_{\rm gh}=-i{\rm ln}[
\frac{{\rm Det}K^{ab}[A]}{{\rm Det}K^{ab}[0]}]=
-i{\rm Tr}[ {\rm ln} K^{ ab}[A] -{\rm ln} K^{ab}[0]]
\label{vach3}
\eea
where $K^{ab}[A]= K^{ab}[A, Q=0]$ is given by eq. (\ref{lghost}).
We can repeat the ghost calculation similar to the gluon
except for the following changes.
There is an over all factor $\frac{1}{4}$
(with $[1+{\rm cosh}(2g\Lambda_j(t))] \rightarrow 2$)
from eq. (\ref{12ff}) because we do not have any
Lorentz matrices in $K^{ab}[A]$ in eq. (\ref{lghost}).
There is another over all factor ($-2$) from eq. (\ref{12ff})
because of the ghost determinant (compare eqs. (\ref{og}) and (\ref{vach3})).
With these two changes we find from eq. (\ref{12ff}) the following expression
for the ghost one-loop effective action
\bea
S^{(1)}_{\rm gh}= -\frac{i}{16 \pi^3} \sum_{j=1}^6
\int_0^\infty \frac{ds}{s} \int d^4x \int d^2p_T
e^{is(p_T^2+i\epsilon)}
[\frac{g\Lambda_j(t)}{{\rm sinh}(sg\Lambda_j(t))}
-\frac{1}{s}].
\label{12ffh}
\eea

\subsection { Non-Perturbative
Gluon Pair Production From Arbitrary $E^a(t)$ via Schwinger Mechanism }

Adding eqs. (\ref{12ff}) and (\ref{12ffh}) for the effective action for gluon and ghost respectively we find
\bea
S^{(1)}_{\rm gl}= \frac{i}{16 \pi^3} \sum_{j=1}^6
\int_0^\infty \frac{ds}{s} \int d^4x \int d^2p_T
e^{is(p_T^2+i\epsilon)}
[g\Lambda_j(t) ~\frac{{\rm cosh}(2sg\Lambda_j(t)}{{\rm sinh}(sg\Lambda_j(t))}
-\frac{1}{s}].
\label{12fft}
\eea
The imaginary part of the above effective action gives real gluon pair
production. The real part of the above equation is infrared divergent as $s \rightarrow \infty$.
However we are interested in the imaginary part of the above effective
action which is not infrared divergent as $s \rightarrow \infty$. This can be
easily checked by making $s \rightarrow is$.
The s-contour integration can be done in the similar way as
was done in \cite{schw,nayak1,nayak2,nayak6}. Using the series expansion
\bea
\frac{1}{{\rm sinh} x}=\frac{1}{x} + 2x \sum_{n=1}^\infty \frac{(-1)^n}{\pi^2 n^2 +x^2}
\eea
we perform the s-contour integration around the pole $s=\frac{in\pi}{|g\Lambda_j(t)|}$ to find
\bea
W=2 {\rm Im} S^{(1)}_{\rm gl}=
\frac{1}{4\pi^3}  \sum_{j=1}^3
\sum_{n=1}^\infty \frac{(-1)^{n+1}}{n} \int d^4x \int d^2p_T |g\Lambda_j(t)|
e^{- \frac{n \pi p_T^2}{|g\Lambda_j(t)|}}.
\label{14ff}
\eea

Hence the probability of non-perturbative gluon (pair)
production per unit time, per unit volume and per unit
transverse momentum from an arbitrary time dependent chromo-electric
field $E^a(t)$ with arbitrary color index $a$=1,2,...8 in SU(3) is given by
\bea
\frac{dW}{dt d^3x d^2p_T}~
=~\frac{1}{4\pi^3} ~~ \sum_{j=1}^3 ~
~|g\Lambda_j(t)|~{\rm ln}[1~+~e^{-\frac{ \pi p_T^2}{|g\Lambda_j(t)|}}],
\label{1new}
\eea
which reproduces eq. (\ref{1}). The expressions for gauge invariant
$\Lambda_j(t)$'s are given in eq. (\ref{lm}).

\section{Conclusion}

To conclude we have studied Schwinger mechanism for gluon pair
production in the presence of an arbitrary time-dependent
chromo-electric background field $E^a(t)$ with arbitrary color index
$a$=1,2,...8 in SU(3). We have obtained an exact result for
the probability of non-perturbative gluon (pair) production
per unit time per unit volume per unit transverse momentum
$\frac{dW}{d^4xd^2p_T}$ from arbitrary $E^a(t)$ by directly
evaluating the path integral.
We have found that the tadpole (or single gluon) effective action does
not contribute to the non-perturbative gluon pair production rate
$\frac{dW}{d^4x d^2p_T}$.
We have found that the exact result for non-perturbative
gluon pair production
is independent of all the time derivatives $\frac{d^nE^a(t)}{dt^n}$
where $n=1,2,...\infty$ and has the same functional dependence on two
casimir invariants $[E^a(t)E^a(t)]$ and $[d_{abc}E^a(t)E^b(t)E^c(t)]^2$
as the constant chromo-electric field $E^a$ result with the replacement:$E^a \rightarrow E^a(t)$.
This result relies crucially on the validity of the shift
conjecture, which has not yet been established. This result may be relevant to study the
production of a non-perturbative quark-gluon plasma at RHIC and LHC
\cite{qgp1,qgp2,qgp3}.

\acknowledgments

This work was supported in part by the National Science
Foundation, grants PHY-0354776 and PHY-0345822.

\end{document}